\documentclass[12pt]{article}

\usepackage[utf8]{inputenc}   
\usepackage{amsmath, amssymb} 
\usepackage{graphicx}         
\usepackage{hyperref}         
\usepackage{geometry}         
\usepackage{setspace}         
\usepackage{enumitem}         
\usepackage{titlesec}         

\usepackage{algorithm}
\usepackage{algpseudocode}
\titleformat{\section}{\large\bfseries}{\thesection}{1em}{}

\title{AI-Driven Consensus: Modeling Multi-Agent Networks with Long-Range Interactions through path-Laplacian Matrices}
\author{Yusef Ahsini, Belén Reverte, and J. Alberto Conejero \\
\small Instituto Universitario de Matemática Pura y Aplicada. Universitat Politècnica de València \\
\small \texttt{aconejero@upv.es}
}
\DeclareMathOperator{\diag}{diag}
\begin{document}

\maketitle

\begin{abstract}
Extended connectivity in graphs can be analyzed through \emph{k}-path Laplacian matrices, which permit the capture of long-range interactions in various real-world networked systems such as social, transportation, and multi-agent networks.
In this work, we present several alternative methods based on machine learning methods (LSTM, xLSTM, Transformer, XGBoost, and ConvLSTM) to predict the final consensus value based on directed networks (Erd\"os-Renyi, Watts-Strogatz, and Barabási-Albert) and on the initial state.
We highlight how different \emph{k}-hop interactions affect the performance of the tested methods. This framework opens new avenues for analyzing multi-scale diffusion processes in large-scale, complex networks.
\end{abstract}



\section{Introduction}

Consensus algorithms are protocols allowing an agent network to agree on certain quantities of interest. Early work on consensus in social systems was introduced by DeGroot \cite{DeGroot1974}, who modeled individuals iteratively updating their opinions to the average of their neighbors’ opinions. However, it is known that other individuals who are not neighbors are also influential in people's opinions \cite{garofalo2025competition} and behaviour \cite{christakis2009connected,christakis2013social}.
In control theory, consensus gained prominence for coordinating multi-agent systems (e.g., fleets of autonomous vehicles or sensor networks) \cite{li2019survey,amirkhani2022consensus}. How this consensus is reached heavily depends on the initial state and network structure.\medskip

Let $G=(V,E)$ be a network with $V$ as the set of nodes and $E$ as the set of nodes. If $A$ is the adjacency matrix of $G$ and $D$ is its degree matrix (a diagonal matrix with the degree of the node $v_i\in V$ in the position $d_{ii}$), the \textit{Laplacian matrix} is defined as \(L = D - A\).
Laplacian matrices and their spectral properties have been extensively studied due to their interest when studying graph connectivity, spreading, and community detection \cite{Mohar1991, Merris1994}. This facilitates measuring robustness, spectral clustering, and partition of networks.\medskip

While a Laplacian matrix captures direct (one-hop) connections connecting adjacent nodes by an edge, Estrada extended the notion of adjacency to include paths of length \(k\) between nodes
generalized to incorporate longer path-based connections, rather than only immediate neighbors by introducing \(k\)-path Laplacian matrices \cite{estrada2012path}. These matrices are constructed from \(k\)-path adjacency matrixes that count the shortest paths of length \(k\) between any pair of nodes, providing a multi-hop representation of the network connectivity. We recall that the case of $k=1$ returns classical Laplacian matrices. If the multiplicity of the zero eigenvalue of \(L^{k}\) is one, then the network is \(k\)-path connected (analogous to one-hop connectivity for the standard Laplacian matrix). These path-based extensions provide a more global view of connectivity beyond immediate neighbors.\medskip

The introduction of path-based Laplacians permits the incorporation of long-range connections and the study of how they impact dynamic network processes. This connectivity extension is particularly relevant for complex networks such as social or transportation systems, where indirect connections (friends-of-friends in a social network or multi-hop routes in a transportation network) significantly influence network behavior. For example, considering \(k>1\), we can accelerate synchronization and consensus in networks by effectively reducing distances between nodes \cite{estrada2018long-range,estrada2021path}. Additionally, by weighting connections along paths, path-Laplacian matrices generalize a foundational graph-theoretic concept and provide a richer framework to study connectivity and diffusion processes on networks.\medskip

The ability of path-based matrices to capture extended connectivity is particularly relevant for complex networks such as social \cite{park2018strength} or transportation systems \cite{viana2011fast}, where indirect connections significantly influence network behavior.
This alternative enhanced formulation of connectivity can be exploited to improve consensus achieving in multi-agent systems.\medskip

This work analyzes several machine-learning methods for estimating the consensus in networks comprising long-range interactions through $k$-path-Laplacian matrices. We study three different types of networks: Erd\"os-Renyi \cite{van2014random}, Watts-Strogatz \cite{watts1998collective}, and Barabási-Albert \cite{barabasi1999emergence}. We also test several machine learning methods to estimate the consensus finally obtained in these networks as time passes.\medskip

The manuscript is organized as follows: In Section \ref{sec:consensus}, we review some aspects on mechanisms on multi-agent systems modeled as networks. Useful machine learning models to study dynamics on networks, mainly based on recurrent neural networks, are described in Section \ref{sec:ml}. The mathematical formalism of consensus through path-Laplacian matrices for directed and undirected networks is shown in Section \ref{sec:theory}. In Section \ref{sec:nm}, we describe the data generation for process which will be used for training, validation, and testing the models. We describe the performance of the models in Section \ref{sec:results} and we outline the conclusions in Section \ref{sec:conclusions}.

\section{Consensus Mechanisms in Multi-Agent Systems}
\label{sec:consensus}

Multi-agent systems can be modeled through a network in which agents interacting with each other are directly connected by an edge. Connectivity is a crucial property for reaching a consensus in a network \cite{Jadbabaie2003,Moreau2005,Ren2005,amirkhani2022consensus}. These foundational works lay on the idea that graph connectivity (often quantified by properties of the Laplacian matrix $L$) is a crucial condition for consensus. For instance, if the agents’ communication graph has a spanning tree across time, the agents’ states will converge to a common value. Connectivity can also be linked to spectral theory. The second-smallest eigenvalue of \(L\), known as the Fiedler value or algebraic connectivity, relates to how well the graph is connected \cite{Fiedler1973}. A non-zero Fiedler value indicates a graph is connected, and larger values imply more robust connectivity. Due to these properties, Laplacians have found broad applications in network analysis, from measuring robustness to facilitating spectral clustering and partitioning of graphs.
Besides, continuous or, at least, frequent enough connectivity is also required to reach the consensus state. 
For static (fixed-topology) networks, consensus algorithms can be analyzed directly using Laplacian matrices. The standard first-order consensus update is given by

\begin{equation}
\mathbf{x}(t+1) = \mathbf{x}(t) - \epsilon L\,\mathbf{x}(t);\quad \text{(discrete time), or}
\end{equation}
\begin{equation}
\dot{\mathbf{x}}(t) = - \epsilon L\,\mathbf{x}(t);\quad \text{(continuous time)},
\end{equation}
where $L$ is the Laplacian matrix of the network, drives all agent states \(\mathbf{x}\) to the agreement (consensus) on an average initial value when the graph is connected.\medskip

The Fiedler value also provides the convergence rate to consensus; that is, a larger Fiedler eigenvalue yields faster consensus \cite{Olfati2007}. These authors also provided a comprehensive framework for consensus and cooperation in networked multi-agent systems, highlighting the role of the Laplacian’s spectral gap in convergence speed and robustness to noise and delays \cite{Hatano2005,Xiao2004}. As an example, one can reduce polarization by properly introducing the information in a network to maximize the spectral gap \cite{racz2023towards}.\medskip

Beyond one-hop (nearest-neighbor) interactions, recent studies consider whether leveraging multi-hop information can improve consensus speed or resilience. Traditional consensus uses only direct neighbor states; however, nodes in many real networks can also be indirectly influenced by more distant nodes (e.g., a friend-of-a-friend in a social network). Path-based Laplacians have been applied to consensus dynamics to capture such influences. Estrada \cite{estrada2012path} showed that allowing agents to utilize not only their immediate neighbors' states but also the states of nodes at a distance \(k\) (via the \(k\)-path Laplacian) can accelerate agreement, see also \cite{Estrada2017,estrada2018long-range,estrada2021path}.\medskip

Long-range interactions provide additional communication "shortcuts" in the network, speeding up information spread and consensus formation. This concept has been extended in control protocols: Ma et al. proposed consensus algorithms based on a transformed \(k\)-path Laplacian and demonstrated improved convergence in networks where agents consider multi-hop neighbors in their updates \cite{ma2020consensus}. Such approaches are particularly relevant in systems where influence extends beyond immediate contacts. In opinion dynamics, models incorporating second- or third-neighborhood influences (sometimes called "social power" or indirect influence) help explain how global consensus or polarization can emerge \cite{Proskurnikov2016}. These insights inform the design of robust coordination strategies in applications ranging from coordinating autonomous vehicles and sensor networks to spreading information or agreement in social networks. Therefore, as network connectivity drives consensus, it is natural to consider how modern data-driven methods can further capture and predict such complex dynamics.\medskip

\section{Machine learning models for studying dynamics on networks}
\label{sec:ml}

Modeling the dynamics of processes on networks often requires capturing both temporal patterns and network structure. In the past decade, researchers have increasingly turned to machine learning—and particularly deep learning models—to model and predict network dynamics. Recurrent Neural Networks (RNNs), especially Long Short-Term Memory (LSTM) networks \cite{hochreiter1997long}, have shown notable success in forecasting traffic flow and travel speeds by effectively learning long-range temporal dependencies \cite{polson2017deep}. Similarly, LSTMs have been applied to social networks to predict the evolution of user engagement and information diffusion over time, leveraging their strength in sequence modeling to anticipate bursts of activity or link formations. Complementary methods such as eXtreme Gradient Boosting (e.g., XGBoost \cite{chen2016xbgoost}) that consist of the use of ensembles of decision trees to model nonlinear relationships have been applied either as standalone models or in hybrid schemes with LSTMs to refine predictions.\medskip

More recently, the transformer architecture \cite{vaswani2017attention} has revolutionized sequence modeling by leveraging self-attention to capture both local and global interactions—yielding improved performance in traffic forecasting and social interaction modeling \cite{lan2021traffic}. Transformers rely on self-attention mechanisms to capture relationships in sequential data, either temporal patterns or correlations between locations, and they are highly parallelizable, enabling very large models. 
In social network analysis, transformers (and their variant, graph transformers) have been employed to model sequences of interactions or events (like sequences of posts or the evolution of relationships), benefiting from attention to identify critical earlier interactions that influence later outcomes \cite{kreuzer2021rethinking}.\medskip

Hybrid approaches combine spatial feature extraction with temporal dynamics, including Convolutional Transformers \cite{khan2023survey,firbas2023characterization} and Convolutional LSTM networks \cite{Shi2015,zha2022forecasting,lozano2021open}. The idea behind them is to use convolutional layers to capture local spatial structure (for example, nearby roads on a map or local clusters in a network) and Transformer/LSTM layers to capture broader contextual or temporal patterns. 
Variants like Graph-structured LSTMs (applying ConvLSTM ideas to graph adjacency instead of image grids) have been developed to handle network data more directly \cite{zayats2018conversation}.
Similarly, for social networks, one could imagine using CNNs on graph-structured data (via graph convolution operations) combined with Transformers for temporal sequences to model phenomena like epidemic spreading or information cascades with high fidelity \cite{hechtlinger2017generalization}.

Finally, emerging models such as Extended LSTM (xLSTM) \cite{beck2023xLSTM} push the limits of recurrent architectures by introducing modifications such as exponential memory gating or parallelizable memory units. Early experiments in sequence modeling (primarily language modeling benchmarks) suggest that, with proper scaling and gating enhancements, xLSTMs can achieve competitive performance to transformers \cite{beck2023xLSTM} without the training difficulties that vanilla LSTMs face. Robust synthetic data generation becomes indispensable to effectively evaluate and validate these machine learning methods, as detailed in the following section.\medskip


\section{Mathematical Formalism of Consensus through Path-Laplacian Matrices}
\label{sec:theory} 

Consider a simple undirected network $G=(V,E)$ with $n=|V|$ nodes (vertices) and $m=|E|$ edges. Two nodes $i,j\in V$ are said to be \emph{adjacent} if an edge $(i,j)\in E$ is connecting them. The \emph{adjacency matrix} $A$ of $G$ is the $n\times n$ matrix with entries $A_{ij}=1$ if $(i,j)\in E$ (and $A_{ij}=0$ otherwise). Let $\deg(i)$ denote the degree of node $i$ (number of neighbors). The diagonal \emph{degree matrix} is $D=\diag(\deg(1),\dots,\deg(n))$. We recall that the \emph{Laplacian matrix} is defined as 
\begin{equation}
L \;=\; D \;-\; A,
\end{equation}
which is a symmetric positive semi-definite matrix of size $n\times n$ that has real, non-negative eigenvalues $0=\lambda_1\le \lambda_2\le \cdots\le \lambda_n$ \cite{godsil2001,chung1996}. 
When modeling, we assume that a node represents an agent, and its adjacent nodes will be connected to it with edges. 
Classical consensus models assume that each agent is influenced only by its nearest neighbors (direct connections). In what follows, we introduce a generalized Laplacian formalism that allows incorporating longer-range interactions beyond immediate neighbors. This generalization is based on graph paths and will lead to an extended family of Laplacian matrices capturing $k$-hop relationships in the network.\medskip 

\subsection{Path-Based Laplacian Matrices}\label{subsec:pathlaplacian} 
A \emph{path} of length $k$ in $G$ is a sequence of $k+1$ distinct vertices $(v_{i_0}, v_{i_1}, \dots, v_{i_k})$ such that each consecutive pair $(v_{i_{r-1}}, v_{i_r})$ is an edge in $E$. The \emph{distance} $d(i,j)$ between two nodes $i$ and $j$ is defined as the length of the shortest path connecting $i$ to $j$. Let $d_{\max}$ denote the graph's diameter (the maximum distance over all node pairs). For each integer $k$ with $1 \le k \le d_{\max}$, we define the \textit{$k$-path adjacency matrix}, denoted $P_k$, as the $n\times n$ (symmetric) matrix with entries 
\begin{equation}
(P_k)_{ij} \;=\;
\begin{cases}
1, & \text{ if} d(i,j) = k,\\
0, & \text{otherwise}.
\end{cases}
\end{equation}
In other words, $P_k(i,j)=1$ if and only if there exists at least one shortest length path $k$ between node $i$ and node $j$. This matrix generalizes the ordinary adjacency matrix ($P_1 = A$) to connections via longer paths. We also define the \textit{$k$-path degree} of node $i$ as
\begin{equation}
\delta_k(i) \;=\; \sum_{j=1}^n (P_k)_{ij}.
\end{equation}
i.e., the number of nodes at distance exactly $k$ from $i$ in the network \cite{estrada2012path}. By construction, $\delta_1(i)=\deg(i)$ is the usual degree. We introduce the \textit{$k$-path Laplacian matrix} $L_k$ of the network $G$ as follows \cite{estrada2012path}: 

\begin{equation}\label{eq:Lk_def} 
(L_k)_{ij} \;=\;
\begin{cases}
-1, & \text{if } d(i,j) = k,\\
\delta_k(i), & \text{if } i=j,\\
0, & \text{otherwise}.
\end{cases}
\end{equation}

Equivalently, we can write $L_k = D_k - P_k$, where $D_k = \diag(\delta_k(1), \dots, \delta_k(n))$ is the diagonal matrix of $k$-path degrees. For $k=1$, $L_1 = D_1 - P_1 = D - A$, which is exactly the classical Laplacian $L$. Each $L_k$ is symmetric and positive semi-definite. In fact, one can verify that for any real vector $y\in\mathbb{R}^n$,
\[
y^\top L_k\,y
\;=\;
\frac{1}{2}
\sum_{\substack{i,j\in V\\ d(i,j)=k}}
\bigl(y_i \;-\; y_j\bigr)^2
\;\ge\; 0.
\]
confirming that all eigenvalues of $L_k$ are non-negative. Moreover, $L_k$ shares the zero-eigenvalue property of the standard Laplacian: since each off-diagonal $-1$ in row $i$ of $L_k$ is balanced by the contribution to the diagonal entry $\delta_k(i)$, it follows that $L_k \mathbf{1} = 0$. Thus $0$ is an eigenvalue of $L_k$ with eigenvector $\mathbf{1}=[1,1,\ldots,1]^T$ for every $k$. The multiplicity of this zero eigenvalue reflects the number of disconnected components in the graph when considering only $k$-distance connections. In particular, if the network $G$ is such that any node can reach any other node through a sequence of hops of length $\le k$ (sometimes said to be $k$-path connected', then $L_k$ has a unique zero eigenvalue. In general, $m$ zero eigenvalues of $L_k$ indicate that the graph can be partitioned into $m$ disjoint subgraphs that are not linked by any $k$-length path. This is a natural generalization of the connectivity interpretation of the ordinary Laplacian ($L_1$). The family ${L_k: 1 \le k \le d_{\max}}$ thus provides a spectrum of Laplacian-like operators, each capturing network connectivity at a different path-length scale. Importantly, higher values of $k$ account for more ``long-range'' links in the network, and as we will see, incorporating such long-range interactions can significantly impact dynamic processes like consensus.\medskip

\subsection{Consensus Dynamics on Undirected Networks}\label{subsec:consensus}
We start reviewing the standard consensus process on undirected network and then describe its extension using the multi-hop Laplacians introduced above. Consider an undirected network of $n$ agents (nodes) whose goal is to reach agreement on some vector $\boldsymbol{\phi}_0=[\phi_{01},\ldots,\phi_{0n}]^T\in \mathbb{R}^n$. Let $\phi_i(t)$ denote the state (value of $\phi$) held by agent $i$ at time $t$. In a consensus algorithm, each agent updates its state by interacting with its neighbors, gradually reducing differences between agents’ values. We distinguish two common frameworks: continuous-time and discrete-time consensus dynamics. Later, we will see how to introduce multi-hop interactions.\medskip

\subsubsection{Continuous-time consensus} 
In the first-order continuous-time model, each agent $i$ adjusts its state with a rate proportional to the discrepancies with its neighbors’ states. This is described by the following system of ordinary differential equations
\begin{equation}
\dot{\phi}_i(t) = \sum_{j \in V_i} \left( \phi_j(t) - \phi_i(t) \right),
\end{equation}
where $V_i$ denotes the set of neighbors of node $i$ (all $1\le j\le n$, $i\ne j$, such that $(i,j)\in E$). Equivalently, in vector form, one can write 
\begin{equation}\label{eq:ct_consensus}
\frac{d\boldsymbol{\phi}(t)}{dt}
\;=\;
-\,
L\,
\boldsymbol{\phi}(t).
\end{equation}
where $\boldsymbol{\phi}(t)=[\phi_1(t),\dots,\phi_n(t)]^\top$ and $L$ is the Laplacian of the network \cite{Olfati2007}. Equation \eqref{eq:ct_consensus} is a linear system whose solution is $\boldsymbol{\phi}(t) =e^{-Lt}\cdot \boldsymbol{\phi}(0)$. For a connected graph, $L$ has a single zero eigenvalue and all other eigenvalues $\lambda_2,\dots,\lambda_n > 0$. In this case, the system \eqref{eq:ct_consensus} drives the agents to a consensus: $\phi_i(t)$ converges to $\phi_{0i}$ as $t\to\infty$ for all $i$. Indeed, one can show $\lim_{t\to\infty} \boldsymbol{\phi}(t) = c \cdot \mathbf{1}$, where $c$ is a constant equal to the average of the initial states $\phi_i(0)$ \cite{Olfati2007,mesbahi2010}. This average is an invariant of \eqref{eq:ct_consensus} because $\mathbf{1}^\top L = \mathbf{0}$, so $\frac{d}{dt}\sum_i \phi_i(t) = \mathbf{1}^\top \dot{\boldsymbol{\phi}}(t) = -\mathbf{1}^\top L \boldsymbol{\phi}(t) = \mathbf{0}$. The eigenvalues of $L$ govern the convergence rate of continuous consensus; in particular, the algebraic connectivity) determines the slowest decay mode and hence the asymptotic convergence speed \cite{Mohar1991,Olfati2007}.\medskip

\subsubsection{Discrete-time consensus} 
In discrete time, consensus is achieved by iterative averaging. At each time step, $t \to t+1$, every agent $v_i$ updates its state as a convex combination of its own and its neighbors’ previous states ($v\in V_i$). A common discrete model can be formulated as
\begin{equation}
\phi_i(t+1) = \phi_i(t) + \varepsilon \sum_{j \in V_i} \left( \phi_j(t) - \phi_i(t) \right),
\end{equation}
where $0 < \varepsilon < 1$ is a small step size, also interpreted as an interaction weight \cite{Olfati2007}. In matrix form, this update can be expressed as
\begin{equation}\label{eq:dt_consensus} 
\boldsymbol{\phi}(t+1)=\bigl(I \;-\; \varepsilon\,L\bigr)\,\boldsymbol{\phi}(t),
\end{equation}
where $I$ is the identity of order $n$. Let $W = I - \varepsilon L$ denote the update's weight matrix. Other generalizations of diffusio that include subdiffusion and super diffusion in networks were stated in \cite{diaz2022time}.\medskip

By construction, each row of \(W\) sums to 1 and \(W_{ij}>0\) if \(j\in V_i\) (with \(W_{ii}>0\) as well). If $0 < \varepsilon < 1/\deg_{\max}$, where $\deg_{\max}$ is the maximum degree in the network, then $W$ has all nonnegative entries and no entry exceeds 1, meaning that $W$ is a stochastic matrix. In such a matrix, only the rows are normalized to sum to 1. On the one hand, if \(W\) is \emph{doubly-stochastic}, that is, both its rows and its columns sum to 1, then the average of the states is invariant under the update, that is 
$\sum_{i=1}^n \phi_i(t+1) =\sum_{i=1}^n \phi_i(t)$ for all $t\ge 0$ and the consensus is reached at $\boldsymbol{\phi_0}$ with $\phi_{01}=\ldots=\phi_{0n}=c$ with $c$ equals the initial average $\frac{1}{n}\sum_{i=1}^n \phi_i(0)$. On the other hand, if \(W\) is only row-stochastic, then the limiting consensus state will be a weighted average determined by the left eigenvector of \(W\) corresponding to eigenvalue 1 \cite{xu2022}. The convergence of \eqref{eq:dt_consensus} is again related to the spectrum of $L$ analogously to the continuous case.
\medskip

\subsubsection {Multi-hop consensus dynamics} 

The previous models can be extended to incorporate \emph{long-range interactions} using the $k$-path Laplacians $L_k$. To model such phenomena, we introduce additional coupling terms that account for neighbors at distances $2,3$, etc. Suppose we fix an integer $k_0\ge 1$ and assign a weight coefficient $\alpha_k \ge 0$ to interactions via $k$-hop neighbors (where $k=1,2,\dots,k_0$). It is natural to assume $\alpha_1 \ge \alpha_2 \ge \cdots \ge \alpha_{k_0}$, reflecting that closer neighbors exert a stronger influence than far neighbors. In some cases, the weights are considered as $\alpha_k = e^{-\beta k}$, 
with \(\beta > 0\), representing an exponential decay of the influence concerning distance. These transformations do not qualitatively transform the diffusion dynamics, i.e., for any parameter, the dynamics remain standard diffusion. However, the Mellin transformation (power law) transforms from normal diffusion to superdiffusion for certain values of the exponent \cite{Estrada2017,estrada2018path}.
In a continuous-time setting, a generalized consensus dynamics can be written as 
\begin{equation}\label{eq:ct_multihop} 
\frac{d\boldsymbol{\phi}}{dt}(t) = -\sum_{k=1}^{k_0} \alpha_k L_k\boldsymbol{\phi}(t). 
\end{equation}

We recall that the standard nearest-neighbor model is obtained from \eqref{eq:ct_consensus}
when $k_0=1$ and $\alpha_1=1$). The matrix $L_{\text{tot}}:= \sum_{k=1}^{k_0} \alpha_k L_k$ from \eqref{eq:ct_multihop} acts as an effective Laplacian matrix for the multi-hop coupling. Like each $L_k$, the combined matrix $L_{\text{tot}}$ is symmetric, positive, and semi-definite, and $L_{\text{tot}}\mathbf{1}=0$. 
Analogously, a multi-hop consensus update can be formulated for the discrete case shown in \eqref{eq:dt_consensus}: 
\begin{equation}\label{eq:dt_multihop} \boldsymbol{\phi}(t+1) = P_c \boldsymbol{\phi}(t),
\end{equation} 
where $P_c$ denotes the combined update matrix $I - \varepsilon \sum_{k=1}^{k_0} \alpha_k L_k$, which generalizes the Perron matrix of the standard model. As before, for $P_c$ to define a proper consensus update, it should have nonnegative entries, and each row should sum to 1. Sufficient conditions can be derived from the step-size $\varepsilon$ and the network’s $k$-path degrees. For instance, if we take $\alpha_k = 1$ for all $k$, a suitable choice of the step size would be
\begin{equation}
0 
\;<\; 
\varepsilon
\;<\;
\frac{1}{\max_{i \in V}\;
\sum_{k=1}^{k_0}\delta_k(i)}.
\end{equation}

\subsection{Consensus Dynamics on Directed Networks}
Suppose that $G=(V,E)$ is now a directed network with $A$ as an adjacency matrix. For each node $v_i\in V$, the \emph{out-degree} of $v_i$ is defined as
$d^{\text{out}}_i = \sum_{j=1}^n A_{ij}$ and the \emph{in-degree} as
$d^{\text{in}}_i = \sum_{j} A_{ji}$. Then, the \emph{directed Laplacian} is given by
\begin{equation}
L^{\text{out}} = D^{\text{out}} - A,
\end{equation}
where \(D^{\text{out}} = \operatorname{diag}(d^{\text{out}}_1,\dots,d^{\text{out}}_n)\). Note that \(L^{\text{out}}\) satisfies $L^{\text{out}} \mathbf{1} = 0$, but, in general, it is not symmetric, so its spectral properties differ from those of the undirected case \cite{Wang2021}. We recall that a directed network is \emph{weight-balanced} if $d^{\text{in}}_i = d^{\text{out}}_i$ for all $1\le i\le n$. In this case, we can show that $W^{-1}L^{\text{out}}W$ is symmetric. Consequently, \(L^{\text{out}}\) has real eigenvalues and a simple zero eigenvalue with the eigenvector \(\mathbf{1}\) (on both left and right) \cite{Park2023}. In the continuous-time case, this guarantees that the total sum of states remains invariant, so the final consensus state is exactly the arithmetic mean of the initial conditions. In contrast, if the network is not balanced, the consensus state will be a weighted average determined by the stationary distribution of the row-stochastic matrix \(W = I - \varepsilon L^{\text{out}}\) \cite{xu2022}.
The discrete-time case is similar to the continuous one. The case of multi-hop interactions is expressed through the \(k\)-hop Laplacian defined as
\begin{equation}
L_k^{\text{out}} = D_k^{\text{out}} - A^{(k)},
\end{equation}
with \(D_k^{\text{out}}\) the diagonal matrix of \(k\)-hop out-degrees and $A^{(k)}$.

\section{Synthetic data generation for network-based analyses}
\label{sec:nm}

Synthetic data generation is essential for developing and testing network algorithms under controlled conditions. Traditional random graph models have long generated synthetic networks with controlled properties. The Erdős–Rényi model produces random graphs by independently connecting nodes with a fixed probability, yielding networks that serve as null models for graph algorithms \cite{erdos1959random}. Watts and Strogatz's small-world networks interpolate between regular lattices and random graphs, capturing the high clustering and short path lengths already observed in real-life examples \cite{watts1998collective}. Barabási and Albert's scale-free model introduced preferential attachment, producing graphs with a power-law degree distribution capturing nature's essence in forming networks \cite{barabasi1999emergence}. These foundational models allow researchers to synthetically create social or transportation network topologies with desired global properties (randomness, small-world characteristics, heavy-tailed connectivity), which are helpful in testing network behavior theories and algorithms' robustness (e.g., community detection or consensus). It is worth mentioning that other techniques, such as Stochastic Block Models (SBMs) \cite{Holland1983} and the Lancichinetti–Fortunato–Radicchi (LFR) \cite{Lancichinetti2008}, 
can capture community structure and realistic degree heterogeneity. In addition, deep generative models enable us to improve synthetic graph generation by incorporating structural patterns from a set of training graphs (e.g., real social networks or road networks) as GraphRNN \cite{you2018graphrnn}, NetGAN \cite{bojchevski2018netgan}, or GraphVAE \cite{simonovsky2018graphvae} do.
However, we will restrict ourselves to the aforementioned three seminal network models in this work.\medskip

Below, we present two distinct procedures for synthetic data generation based on the consensus dynamics outlined previously. The first scenario considers a \emph{local (one-hop)} interaction model, while the second scenario extends interactions through an \emph{exponential (multi-hop)} model. Both approaches begin to generate a network and are subsequently converted into directed networks by assigning orientations to edges, allowing bidirectional connections with a given probability.\medskip

\subsection{Base Case: Local (One-Hop) Interactions}
\label{subsec:base_case_data}

In the local interaction scenario, the synthetic network generation process begins by sampling an undirected random graph $\tilde{G}$ with $n$ nodes according to the chosen random model (Erd\"os-Renyi, Watts-Strogatz, or Barabàsi-Albert). Each undirected network acts as the structural backbone for the subsequent directed network construction. 

Let $G=(V,E)$ an undirected network and $G'=(V',E')$ the new directed network. Here $V=V'$. We assign a direction to each edge $(u,v)\in E$, creating an initial directed edge $(u,v)\in E'$. Additionally, the reverse edge ($(v,u)$) is introduced probabilistically (e.g., with probability $0.3$), thus resulting in a directed network $G'$. The rest of the calculations are performed over $G'$. We compute its adjacency matrix $A$ and the diagonal out-degree matrix $D^{\mathrm{out}}$, from which the directed Laplacian matrix is derived as
\begin{equation}
L = D^{\mathrm{out}} - A.
\end{equation}
Consistent with the previously described consensus dynamics, we introduce the auxiliary matrix $K = L + A$ and determine the iterative update matrix $P$ as:
\begin{equation}
P = I - \epsilon L,\quad \text{where}\quad \epsilon = \frac{1}{c\,\max(K)},
\end{equation} with $c$ typically chosen around $100$ to ensure numerical stability. The initial state vector $\phi(0)$ is constructed from the diagonal elements of $K$ (i.e., $\phi_i(0)=K_{ii}$). The consensus evolution then iterates according to:
\begin{equation}
\phi(t+1) = P\,\phi(t),
\end{equation}
until the state convergence reaches a predefined tolerance $\tau$ or until the maximum allowed iterations are performed. Only the first ten state vectors and the final convergent mean state are retained to ensure computational efficiency and manageable data volume.

\subsection{Exponential Case: Extended (Multi-Hop) Interactions}
\label{subsec:exponential_case_data}

The exponential (multi-hop) interaction scenario generalizes the one-hop procedure by explicitly considering higher-order node interactions. The network diameter $d_{\max}$ of $G'$ (defined on the underlying undirected topology) is calculated, determining the maximum interaction range. For each interaction distance $k$ from $1$ to $d_{\max}$, the corresponding $k$-hop adjacency matrix $A^{(k)}$ is computed through repeated matrix multiplication. Each $A^{(k)}$ leads to the associated $k$-path Laplacian matrix:
\begin{equation}
L_k = D_k - A^{(k)},
\end{equation}
where $D_k$ is a diagonal matrix containing the out-degree sums of $A^{(k)}$.  The interactions across longer paths are weighted by an exponentially decaying factor $e^{-\alpha k}$, where $\alpha > 0$, which controls the influence decay rate. Thus, the multi-hop consensus iterative update matrix $P_{\text{multi}}$ is defined as:
\begin{equation}
P_{\text{multi}} = I - \epsilon \sum_{k=1}^{d_{\max}} e^{-\alpha k} L_k,
\end{equation}
where the parameter $\epsilon$ is carefully chosen to maintain nonnegative matrix entries and row normalization, ensuring stochastic-like behavior.

The iterative consensus evolution begins from an initial state $\phi(0)$ constructed from the diagonal entries of the one-hop matrix $K = L + A$. The state vector is then updated iteratively according to:
\begin{equation}
\phi(t+1) = P_{\text{multi}}\,\phi(t),
\end{equation}
and the procedure continues until convergence (within tolerance $\tau$) or the maximum iteration limit is reached. As before, only the initial ten states and the final mean consensus state are recorded, ensuring computational tractability while capturing essential dynamic features.

\subsection{Pseudocode of the Data Generation}
\label{subsec:datagen_pseudocode}

Algorithm~\ref{alg:pseudocode} summarizes the synthetic data generation steps for local and exponential consensus cases, explicitly detailing graph construction, edge orientation, Laplacian calculations, iterative consensus procedures, and data storage considerations.

\begin{algorithm}[ht!]
\caption{Synthetic Data Generation with Base (One-Hop) and Exponential (Multi-Hop) Consensus}
\label{alg:pseudocode}
\begin{algorithmic}[1]
\Require Number of nodes $n$, Random graph model $\mathcal{M}\in\{\text{BA},\text{WS},\text{ER}\}$, Case type $\in\{\texttt{base},\texttt{exponential}\}$, Bidirectional probability $p_b$, Tolerance $\tau$, Maximum iterations $\mathit{iter}_{\max}$, Decay parameter $\alpha$ (for multi-hop), Record length $T=10$
\Ensure Time-series states $\{\phi(0),\dots,\phi(T-1)\}$, Final consensus value $\frac{1}{n}\sum_{i=1}^{n}\phi_i(\infty)$

\State Generate random undirected graph $\tilde{G}$ (BA, WS, or ER).
\State Create directed graph $G$ from $\tilde{G}$, orienting edges and adding reverse edges with probability $p_b$.
\State Compute adjacency matrix $A$ and out-degree diagonal matrix $D^{\mathrm{out}}$ from $G$.
\State Form directed Laplacian $L = D^{\mathrm{out}} - A$; set $K = L + A$, $\epsilon = 1/(100\cdot\max(K))$.

\If{\texttt{case type} = \texttt{base}}
    \State Set $P = I - \epsilon\,L$.
\Else
    \State Compute $d_{\max} = \text{diameter}(\tilde{G})$.
    \State Calculate matrices $A^k$ and Laplacians $L_k$ for $1\le k\le d_{\max}$.
    \State Set $P = I - \epsilon\sum_{k=1}^{d_{\max}} e^{-\alpha k}L_k$.
\EndIf

\State Initialize $\phi(0)=\mathrm{diag}(K)$.
\For{$t=0$ to $\mathit{iter}_{\max}-1$}
    \State $\phi(t+1) = P\,\phi(t)$
    \If{$\|\phi(t+1)-\phi(t)\|_{\infty}\le\tau$}
        \State \textbf{break}
    \EndIf
\EndFor
\State Record $\{\phi(0),\dots,\phi(\min(T-1,t))\}$ and $\frac{1}{n}\sum_{i}\phi_i(t)$.
\end{algorithmic}
\end{algorithm}

\section{Results}
\label{sec:results}

In this section, we present the performance assessment of our proposed models under different directed graph configurations, focusing on accurately predicting the final consensus value while accommodating the extended connectivity provided by $k$-path Laplacian matrices. We generated, for every graph type (Erdős–Rényi, Watts–Strogatz, and Barabási–Albert), node size (25, 50, 100, 200, and 300), and case (Base and Exponential), a fixed-size training set of 2400 samples and a test set of 600 samples. Each model—XGBoost, LSTM, Extended-LSTM, Transformer, and ConvLSTM—was exposed to identical training and test data to allow fair comparisons. Unless otherwise stated, we reserve 20\% of the training data for validation, leading to an 80/20 train–validation split. Al the models and training data are accessible at \url{https://github.com/yusef320/AI-Driven-Consensus}\medskip

We systematically tuned hyperparameters through predefined search grids tailored to each approach. For instance, maximum tree depth, number of estimators, and learning rate were explored for XGBoost; hidden layer sizes, projection sizes, and multiple LSTM layers were considered for the recurrent-based architectures; and varying numbers of heads, hidden dimensions, and layers were tested for the Transformer models. A mean-squared-error objective guided training, combined with early stopping triggered after 30 consecutive epochs (or boosting rounds for XGBoost) without validation loss improvement. Final evaluations were performed on the 600-sample holdout test set. The main performance metrics include Root Mean Squared Error (RMSE) and Mean Absolute Percentage Error (MAPE).\medskip

Post-training, predictive performance was evaluated primarily using the RMSE and the MAPE. RMSE provides a clear representation of the typical magnitude of prediction errors in the original units of the target variable. At the same time, MAPE offers a normalized error measure by expressing the average absolute error as a percentage of the true values.\medskip

In the following subsections, we compare predictive accuracy and computational overhead for each candidate model, emphasizing how long-range interactions (via higher-order path-Laplacian matrices) influence the quality of the consensus estimation. As discussed previously, larger values of $k$ allow greater leverage of indirect connectivity, potentially strengthening information flow in highly sparse or disconnected graphs. The empirical findings highlight the importance of carefully chosen hyperparameters and how each architecture scales in the presence of multi-hop influences.\medskip

\subsection{Validation}

\begin{figure}[H]
    \centering
    \includegraphics[width=1\linewidth]{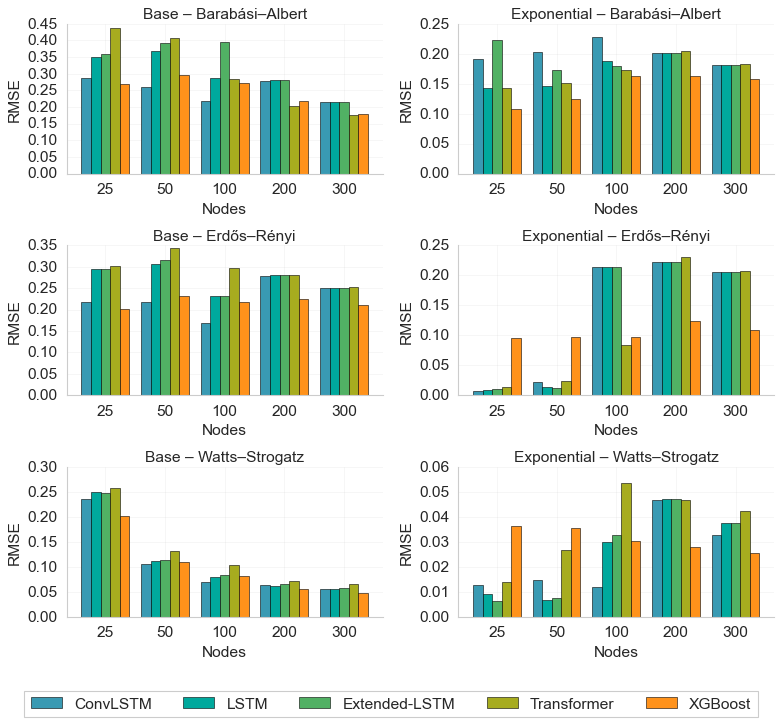}
    \caption{Comparison of RMSE values for five models (ConvLSTM, LSTM, Extended-LSTM, Transformer, XGBoost) in both base and exponential scenarios across three network topologies (Barabási–Albert, Erdős–Rényi, Watts–Strogatz) and varying node sizes.}
    \label{fig:rmse_bar}
\end{figure}

Figure \ref{fig:rmse_bar} presents the RMSE achieved by five predictive models—ConvLSTM, LSTM, Extended-LSTM, Transformer, and XGBoost—for both the base (left column) and exponential (right column) scenarios in three network types: Barabási–Albert (top row), Erdős–Rényi (middle row), and Watts–Strogatz (bottom row). Each chart compares the models’ RMSE for increasing node sizes (25, 50, 100, 200, 300). Overall, the exponential scenario reduces the RMSE relative to the base scenario, suggesting that multi-hop interactions improve predictive accuracy. These patterns highlight how network topology, including long-range interactions and model architecture, each influence consensus prediction accuracy.\medskip

\begin{figure}[H]
    \centering
    \includegraphics[width=1\linewidth]{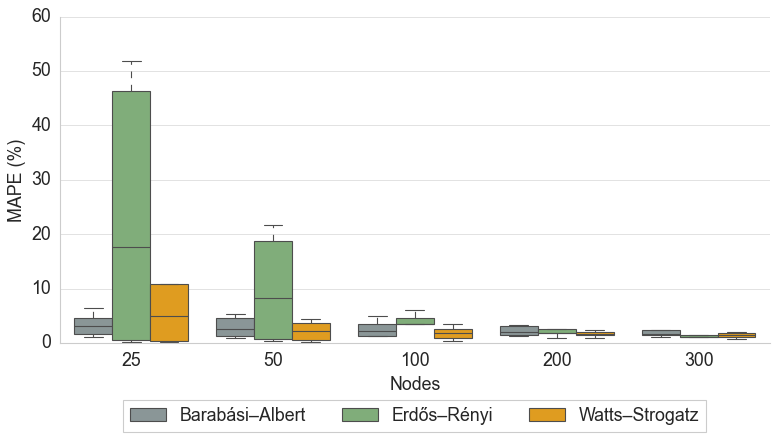}
    \caption{Mean Absolute Percentage Error (MAPE) for consensus predictions across network sizes for Barabási–Albert, Erdős–Rényi, and Watts–Strogatz models.}
    \label{fig:mape_graph}
\end{figure}

In Figure \ref{fig:mape_graph}, we show the Mean Absolute Percentage Error (MAPE) across different network sizes (25, 50, 100, 200, 300 nodes) for three topologies: Barabási–Albert, Erdős–Rényi, and Watts–Strogatz. The bars indicate that at smaller scales (25 and 50 nodes), Erdős–Rényi networks exhibit substantially higher MAPE values, suggesting more variability or unpredictability in their consensus dynamics. Watts–Strogatz networks present intermediate error levels, while Barabási–Albert remains comparatively lower even for smaller node counts. As the number of nodes increases, the MAPE for all three topologies converges to very low values, implying that larger networks become easier to predict regardless of their underlying structure. This pattern highlights the strong influence of network topology—especially the randomness in Erdős–Rényi—on predictability at small scales, whereas scale-free (Barabási–Albert) and small-world (Watts–Strogatz) networks remain more stable in smaller configurations.\medskip

\begin{figure}[H]
    \centering
    \includegraphics[width=1\linewidth]{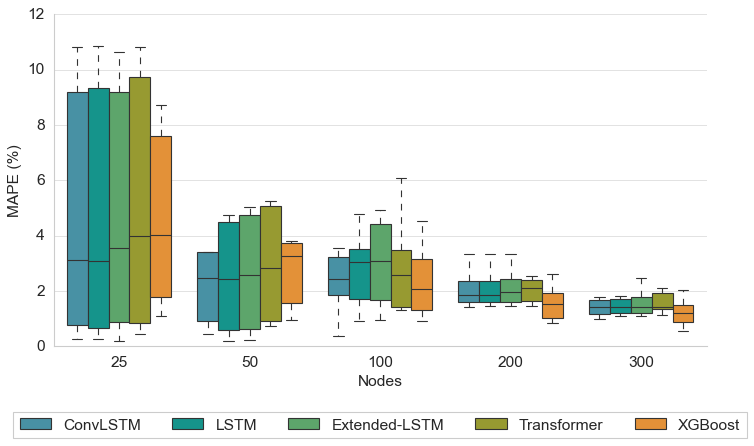}
    \caption{Box-and-whisker plots of MAPE for ConvLSTM, LSTM, Extended-LSTM, Transformer, and XGBoost at different network sizes.}
    \label{fig:mape}
\end{figure}

Figure \ref{fig:mape} displays box-and-whisker plots of the Mean Absolute Percentage Error (MAPE) for five predictive models—ConvLSTM, LSTM, Extended-LSTM, Transformer, and XGBoost—across varying network sizes (25, 50, 100, 200, 300). Each box illustrates the distribution of prediction errors, with taller boxes and whiskers indicating more considerable variability. At small network sizes (25–50 nodes), specific methods (e.g., Extended-LSTM, Transformer, and XGBoost) seem to produce slightly higher median MAPEs than ConvLSTM and LSTM. When you get to 200–300 nodes, the boxes for most methods become narrower and cluster closer together. In other words, as the network size increases, all models tend to converge toward lower MAPE values, reflecting improved stability and predictability. Overall, the results emphasize that both model architecture and mostly network size significantly affect prediction accuracy.\medskip

\subsection{Computational Efficiency}

\begin{figure}[H]
    \centering
    \includegraphics[width=1\linewidth]{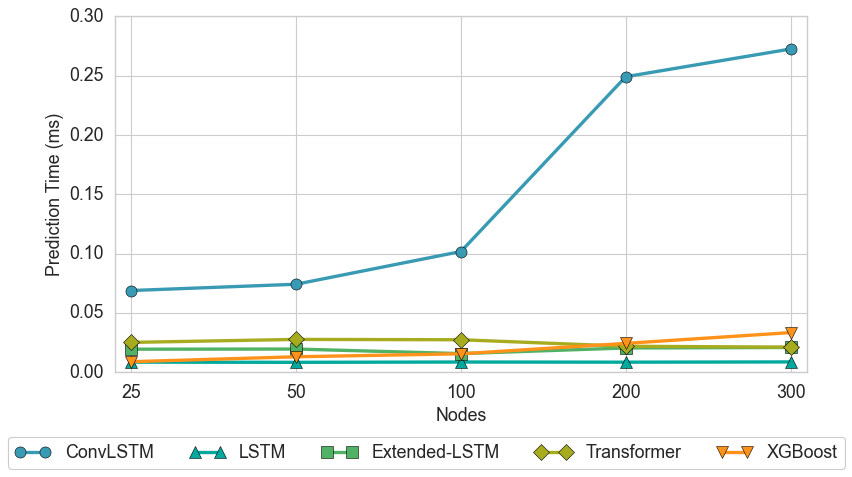}
    \caption{Prediction time comparison for ConvLSTM, LSTM, Extended-LSTM, Transformer, and XGBoost across various network sizes.}
    \label{fig:time}
\end{figure}

Figure \ref{fig:time} illustrates the average prediction time (in milliseconds) for five models—ConvLSTM, LSTM, Extended-LSTM, Transformer, and XGBoost—across networks of increasing size (25, 50, 100, 200, 300 nodes). The graph shows that ConvLSTM’s prediction time scales significantly as the number of nodes grows, surpassing 0.25 ms at 300 nodes. In contrast, LSTM, Extended-LSTM, Transformer, and XGBoost remain relatively stable and low in prediction time, indicating they handle larger networks more efficiently. This highlights the importance of computational overhead when selecting a model for real-time or large-scale scenarios.\medskip

\section{Conclusions}
\label{sec:conclusions}

Our study introduces a novel framework that significantly advances the modeling of consensus dynamics in multi-agent networks by incorporating machine learning methods to include the effect of long-range interactions through path-Laplacian matrices. By extending the classical one-hop Laplacian formulation to account for $k$-hop connections, we have demonstrated that capturing extended connectivity not only estimates the final consensus state, but also enhances the robustness of consensus protocols. Such an approach results in particular interest in directed networks, when the consensus is not just the average of components and depends on the inner structure of the network.\medskip

The experimental analysis—conducted on synthetic networks generated using Erdös-Rényi, Watts-Strogatz, and Barabási-Albert models—underscores the pivotal role of multi-hop interactions. In the base case of local interactions, traditional consensus mechanisms perform adequately; however, the exponential (multi-hop) approach reveals a marked improvement in consensus speed and accuracy. This suggests that considering indirect, long-range influences provides additional communication ``shortcuts,'' thereby effectively reducing the distance between nodes and promoting faster agreement.\medskip

Moreover, our comparative study of various machine learning techniques—including LSTM, xLSTM, Transformer, XGBoost, and ConvLSTM—highlights the considerable potential of data-driven methods in predicting the final consensus state. Notably, these advanced architectures excel in capturing both local and global temporal dynamics inherent in complex networks. Integrating deep learning with advanced graph-theoretic constructs paves the way for more efficient and adaptive consensus algorithms, particularly relevant in applications such as autonomous systems and distributed sensor networks.\medskip

In summary, the most compelling outcomes of our work are the successful extension of the Laplacian framework to encapsulate long-range interactions and the demonstrated efficacy of state-of-the-art machine learning models in predicting consensus dynamics. These insights enrich our theoretical understanding and offer practical avenues for enhancing consensus mechanisms in real-world multi-agent systems.\medskip

\subsection*{Authors contributions}
Conceptualization, Y.A., B.R., J.A.C.; investigation,  Y.A., B.R., J.A.C.; software, Y.A., J.A.C.; validation, Y.A., visualization, Y.A.; formal analysis, B.R., J.A.C.;  writing---original draft preparation, Y.A., B.R., J.A.C.; writing---review and editing, Y.A., B.R., J.A.C.; supervision, J.A.C. All authors have read and agreed to the published version of the manuscript.

\subsection*{Funding}
This research was funded by the European Union - NextGenerationEU, ANDHI project CPP2021-008994 and PID2021-124618NB-C21, by MCIN/AEI/10.13039/501100011033 and by ``ERDF A way of making Europe'', from the European Union.

\subsection*{Data availability}
No data was generated for this work.

\subsection*{Conflict of interest}
The authors declare no conflict of interest.

\bibliographystyle{plain}
\bibliography{references}

\end{document}